\documentclass[]{spie}

\usepackage{amsmath,amsfonts,amssymb}
\usepackage{graphicx}
\usepackage{braket}
\usepackage[colorlinks=true, allcolors=blue]{hyperref}
\usepackage[capitalise]{cleveref}

\newcommand{\mE}{\mathcal{E}}
\newcommand{\mJ}{\mathcal{J}}
\newcommand{\mL}{\mathcal{L}}
\newcommand{\mQ}{\mathcal{Q}}
\newcommand{\mD}{\mathcal{D}}
\newcommand{\mU}{\mathcal{U}}
\newcommand\D{\operatorname{d}\!}
\newcommand{\mQF}{\mathcal{G}}

\title{Non-equilibrium quantum thermodynamics of a particle trapped in a controllable time-varying potential}

\author[a]{Qiongyuan Wu}
\author[a]{Matteo Carlesso}

\affil[a]{Centre for Quantum Materials and Technologies,
School of Mathematics and Physics, Queens University, Belfast BT7 1NN, United Kingdom}

\authorinfo{Further author information: (Send correspondence to M.C.)\\Q.W.: E-mail: qwu03@qub.ac.uk\\M.C.: E-mail: m.carlesso@qub.ac.uk}

\begin{document} 
\maketitle

\begin{abstract}

Non-equilibrium thermodynamics can provide strong advantages when compared to more standard equilibrium situations. 
Here, we present a general framework to study its application to concrete problems, which is valid also beyond the assumption of a Gaussian dynamics.
 We consider two different problems: 1) the dynamics of a levitated nanoparticle undergoing the
transition from an harmonic to a double-well potential; 2) the transfer of a quantum state across a double-well potential through classical and quantum protocols. In both cases, we assume that the system undergoes to decoherence and thermalisation. In case 1), we construct a numerical approach to the problem and study the non-equilibrium thermodynamics of the system. In case 2), we introduce a new figure of merit to quantify the efficiency of a state-transfer protocol and apply it to quantum and classical versions of such protocols.

\end{abstract}


\section{INTRODUCTION}\label{sec:intro}

Over the last decade, optomechanics showed a promising potential in controlling and manipulating mesoscopic systems \cite{Aspelmeyer_2014,Millen_2020,Gonzalez-Ballestero_2021}, whose implementations already have an important impact on quantum technologies \cite{Delic_2020,Magrini_2021,Tebbenjohanns_2021}. On the other hand, quantum thermodynamics provides the energetic footprint of quantum processes, thus characterising their efficiency \cite{Kosloff_2013,Goold_2016,Alicki_2018,Deffner_2019}. The combination of these two burgeoning fields will yield fruitful applications in the field of quantum technologies in the near future \cite{Rossnagel_2014,Brunelli_2018,Ciampini_2021}.\\

The characterization of entropy and its production is among the most important tasks in quantum thermodynamics \cite{Landi_2021}. By connecting the fields of quantum thermodynamics and quantum information, they describe the process from a quantum-information perspective and provide the first step in quantifying the efficiency of the process, thus clearing the path for possible optimizations \cite{Hammam_2021,Ji_2022}. 
Once understood how to characterise the non-equilibrium thermodynamics of an out-of-equilibrium process \cite{Qiongyuan_2022},
we focus on the task of transferring a quantum system from one side to  another of a potential (say one featuring a double-well) while preserving its information content (i.e., its quantum state) \cite{Qiongyuan_2023}. This can be considered a fundamental first step toward the implementation of mesoscopic quantum technologies. To identify the best state-transfer protocol, the quantification of its efficiency becomes pivotal. Several aspects need to be accounted: the fidelity of the final state with respect to a target one, the speed of the protocol, and its thermodynamical irreversibility. \\

Here, we present a general approach to characterise the thermodynamics of a process that undergoes a time-varying potential and the influences of an external environment. We demonstrate our approach by applying it to a specific model. We consider a particle under the action of a harmonic potential turning into a double-well. We characterise the process under the thermodynamical perspective by computing the entropy production and its rates. 
We then introduce a novel figure of merit – namely, the protocol grading -- to quantify the efficiency of the state-transfer protocol in terms of fidelity, speed and thermodynamical irreversibility of the transfer protocol. We apply our approach to a quantum system which is transferred from one to the other well of a double-well potential.

\section{Thermodynamics of a quantum system in a time-varying potential}

\begin{figure} [ht]
    \begin{center}
    %\begin{tabular}{c} %% tabular useful for creating an array of images 
    \includegraphics[width=8cm]{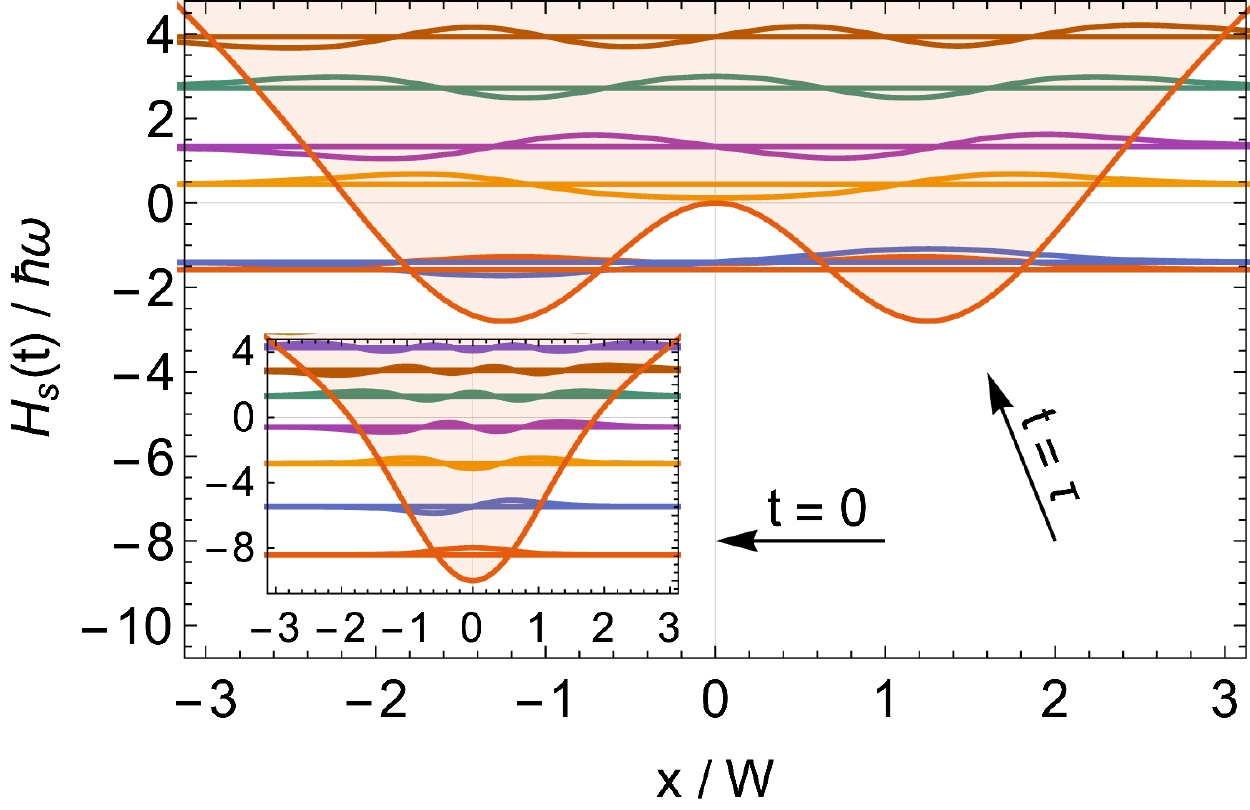}
    %\end{tabular}
    \end{center}
    \caption[example]
    {\label{fig:paper1_potential}
    Time-varying potential (red line) and the instantaneous energies (straight colored lines) and corresponding eigenstates (oscillating colored lines). Figure taken from Ref.~\citenum{Qiongyuan_2022}.}
\end{figure}

To study the thermodynamics of a system under a non-equilibrium transformation, we focus on the example of a quantum harmonic oscillator of mass $m$ and frequency $\omega$ whose potential is transformed to a double-well. The free Hamiltonian reads
\begin{equation}\label{equ:systemhamiltonian2}
    H_\text{s} (t) = \frac{p^2}{2m} + \frac{1}{2}m\omega^2 x^2 + H_\text{add}(t),
\end{equation}
where
\begin{equation}\label{equ:doublepotential}
    H_\text{add}(t) = - {\mE}\left(\alpha(t) + \frac{(1- \alpha(t))}{2}\frac{x^2}{W^2}\right) e^{-\frac{x^2}{2W^2}},
\end{equation}
with $\mE$ and $W$ being a suitably chosen energy scale and length. The potential at time $t=0$ and $t=\tau$ with $\alpha(t)=1-t/\tau$ is shown in \cref{fig:paper1_potential}.

Further, we consider the system in interaction with its surrounding environment, which can be described with the following master equation
\begin{equation}\label{equ:systemdynamics}
    \frac{\D\rho}{\D t} = -\frac{i}{\hbar} [H_\text{s}, \rho] + D_\text{lc}[\rho]+ D_\text{th}[\rho] ,
\end{equation}
where 
\begin{equation}\label{equ:localisationdissipator}
    D_\text{lc}[\rho] = - \Lambda[x,[x,\rho]],
\end{equation}
is a localisation term of coupling strength $\Lambda$, which quantifies the decoherence effects of the collision of the residual gas in the vacuum. On the other hand, we have 
\begin{equation}\label{equ:thermaldissipator}
    D_\text{th}[\rho] = \gamma\left[(\bar{n}+1) {\mL}_{a}(\rho) + \bar{n} {\mL}_{a^\dag}(\rho)\right],
\end{equation}
being the thermal master equation describing the heat exchange between the system and the environment. Here ${\mL}_O(\rho)=O \rho O^\dagger - \lbrace O^\dagger O , \rho \rbrace/2$ and we employed $x=\sqrt{\hbar/2m \omega}(a^\dag+a)$ and $p=i\sqrt{\hbar m \omega/2}(a^\dag-a)$. Such a term is described in terms of $\gamma$ and $\bar n= (e^{\beta \hbar\omega} -1)^{-1}$ being respectively the coupling strength between system and environment and the mean number of excitations at the inverse temperature $\beta$. \\

We will characterise the non-equilibrium thermodynamics of the system through the use of the 
Wehrl entropy $S_\mQ$. For a system of $N=2j+1$ degrees of freedom, the latter reads \cite{Wehrl_1979, Brunelli_2018,Santos_2018}
\begin{equation}\label{equ:wehrlentropy}
    S_\mQ = -\frac{N}{4\pi} \int \D\Omega~\mQ(\Omega) \ln \mQ(\Omega),
\end{equation}
where $\D\Omega=\sin(\theta)\D\theta\D\phi$, while $\mQ(\Omega)=\bra{\Omega}\rho\ket{\Omega}$ is the Husimi-Q function associated with the state $\rho$.
Its time derivative gives the Wehrl entropy production
\begin{equation}\label{equ:decomposewehrlentropy}
    \frac{\D S_\mQ}{\D t}
    =\frac{\D S_\mU}{\D t}+\frac{\D S_{\mD_\text{lc}}}{\D t}+\frac{\D S_{\mD_\text{th}}}{\D t},
\end{equation}
whose terms correspond to those in the master equation in \cref{equ:systemdynamics}.
The terms connected to the dissipators $\mD$ are further decomposed in the irreversible entropy production rate $\Pi$ and the entropy flux rate $\Phi$ as \cite{Santos_2018,Landi_2020}
\begin{equation}
    \frac{\D S_\mD}{\D t} = \Pi - \Phi.
\end{equation}
The second principle of thermodynamics imposes an ever-increasing irreversible entropy production rate $\Pi$, while the whole entropy production rate  of the system $\D S_\mD/\D t$ can be negative for a sufficiently large positive entropy flux rate $\Phi$.\\

\begin{figure} [ht]
    \begin{center}
    %\begin{tabular}{c} %% tabular useful for creating an array of images 
    \includegraphics[width=8cm]{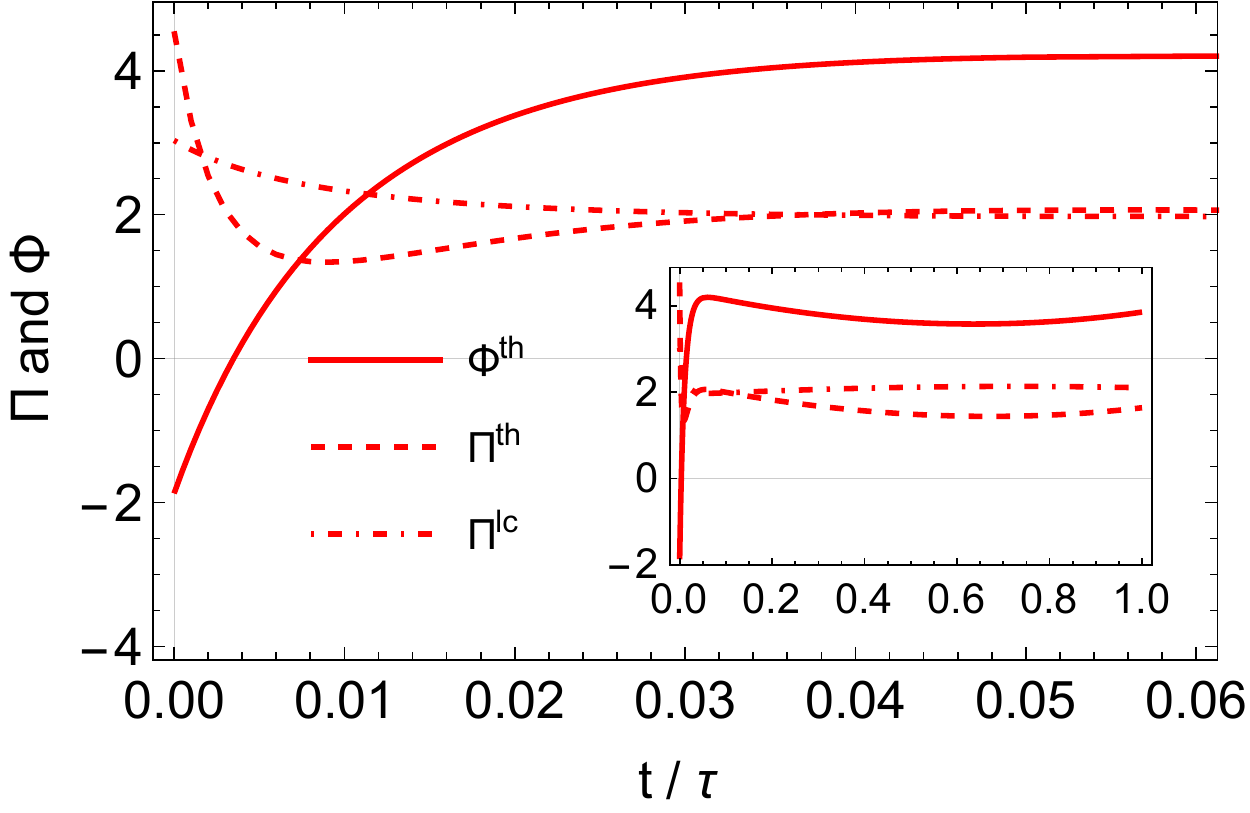}
    %\end{tabular}
    \end{center}
    \caption[example] 
    %>>>> use \label inside caption to get Fig. number with \ref{}
    {\label{fig:paper1_irreversibleentropy}
    Irreversible entropy production rates $\Pi^\text{th}$, $\Pi^\text{lc}$ and the entropy flux rate $\Phi^\text{th}$ for the dynamics at the beginning of the dynamics of a 25-level spin system with a Gaussian-to-double-well potential. Their entire evolution from $t=0$ to $t=\tau$ is illustrated in the inset. Figure taken from Ref.~\citenum{Qiongyuan_2022}.}
\end{figure}

To compute the explicit form of the latter contributions for our model, we suitably discretise the system using the Holstein Primakoff (HP) transformation \cite{Holstein_1940,Gyamfi_2019,Vogl_2020}, which introduces an effective bosonic system with a fixed dimension $N$;
and write it in terms of angular momentum operators $\{J_x, J_y, J_z, J^2\}$. In such a way, the Wehrl entropy can be computed through the following states \cite{Radcliffe_1971,Santos_2017,Santos_2018,Landi_2020}
\begin{equation}
    \ket{\Omega} = e^{- i \phi J_z} e^{-i \theta J_y} \ket{j,j}.
\end{equation}
Here $\ket{j,j}$ is the angular momentum state with largest quantum number of $J_z$ and $\Omega=\{\theta, \phi \}$ is the set of Euler angles identifying the direction along which the coherent state points. Using this method, we obtain the numerical expressions for $\Pi$ and $\Phi$, which correspond to those computed analytically, being
\cite{Santos_2018} 
\begin{equation}\label{equ:localisationentropyproductionrate}
    \Pi^\text{lc} =\Lambda\frac{ N}{4\pi}   \int \D \Omega~\frac{|\mJ_x(\mQ)|^2}{\mQ},\quad\text{and}\quad \Phi^\text{lc}=0,
\end{equation}
for the localisation process, and 
\begin{equation}\label{equ:thermoentropyfluxrate}
    \Phi^\text{th}=\gamma\frac{j(2j+1)}{4\pi} \int \D\Omega \sin\theta \left( \frac{2j \mQ \sin\theta}{(2\overline{n}+1)-\cos\theta} - \partial_\theta \mQ \right),
\end{equation} 
\begin{equation}\label{equ:thermoentropyproductionrate}
    \Pi^\text{th}= {\gamma}\frac{(2j+1)}{8\pi} \int \frac{\D \Omega}{\mQ} \left\{ 
    |\mJ_z(\mQ)|^2 \frac{\large[(2\overline{n}+1)\cos\theta-1\large]}{\tan\theta\sin\theta}
    +\frac{[2 j \mQ \sin\theta {+} (\cos\theta{-} (2\overline{n}+1))\partial_\theta \mQ]^2}{(2\overline{n}+1)-\cos\theta} \right\},
\end{equation}
for the thermalisation process. Their evolution is shown in \cref{fig:paper1_irreversibleentropy}, where they approach a non-equilibrium steady state. The latter is characterised by the relation
\begin{equation}\label{equ:ratesrelation}
    \Pi^\text{th}+\Pi^\text{lc}-\Phi^\text{th}=0,
\end{equation}
where none of the quantities is individually zero.

\section{State transfer protocols and quantification of their efficiency}

Owning the methods to study the thermodynamics of a system undergoing a non-equilibrium transformation, we are now  able to discuss the second problem at hand: the quantification of the efficiency of a state-transfer protocol. We thus introduce a new figure of merit -- namely the protocol grading $\mQF$ -- specifically designed to emphasise a protocol that gives  faithful results (high fidelity), is fast (near to the quantum speed limit) and does not have a strong entropic cost (small production of irreversible entropy).\\

The idea behind the construction of the protocol grading $\mQF$ is the following.
We want a quantity, bounded between 0 and 1, that goes towards 1 if the protocol performs well (high fidelity, near to the quantum speed limit, and with a small production of irreversibility), and 0 if not.
We explicitly account the quantum speed limit $g_\text{\tiny S}$, the experimental quality $g_\text{\tiny Q}$, and the thermodynamic cost $g_\text{\tiny T}$ of the protocol, and construct $\mQF$ as \cite{Qiongyuan_2023}
\begin{equation}\label{equ:quantifier}
    \mQF=g_\text{\tiny S}\times g_\text{\tiny Q} \times g_\text{\tiny T}.
\end{equation}
Here, we  quantify the speed $g_\text{\tiny S}$ with the following coarse-grained function,
\begin{equation}\label{eq.def.gs}
    g_\text{\tiny S} = \max \left\{ 0,\, \left(1-0.1\times\log_{10}\frac{\tau}{\tau_\text{\tiny QSL}}\right) \right\},
\end{equation}
where $\tau$ is the time of the protocol, and $\tau_\text{\tiny QSL}$ is a fundamental lower bound to the time required to distinguish two states during an evolution, which is determined by the quantum speed limit \cite{Deffner_2017}. 

The experimental quality $g_\text{\tiny Q}$ is identified as
\begin{equation}
    g_\text{\tiny Q} = F_\text{exp}(\rho_f,\rho_\text{\tiny TG}), 
\end{equation}
where $F$ is the fidelity between the final state of the system at the end of the protocol $\rho_f$ and the target state $\rho_\text{\tiny TG}$, being  the state of the system as if it would evolve in the target well already from $t=0$. 

Finally, to quantify the thermodynamic cost $ g_\text{\tiny T}$, we use
\begin{equation}\label{eq.def.gt}
      g_\text{\tiny T} = e^{-\Sigma_\text{ir}},
\end{equation} 
which resembles the expression from fluctuation theorems \cite{Crooks_1999,Wang_2002,Jarzynski_2004,Funo_2018}, and it favors the process with small irreversible entropy production. The explicit form of $\Sigma_\text{ir}$, which is  the total irreversible entropy producted during the process, is given by 
\begin{equation}\label{equ:ratetoproduction}
      \Sigma_\text{ir} = \int_0^\tau \D t\,\Pi(t),
\end{equation}
with $\Pi$ that can be computed via \cref{equ:localisationentropyproductionrate} and \cref{equ:thermoentropyproductionrate}.\\

\begin{figure} [ht]
    \begin{center}
    %\begin{tabular}{c} %% tabular useful for creating an array of images 
    \includegraphics[width=8cm]{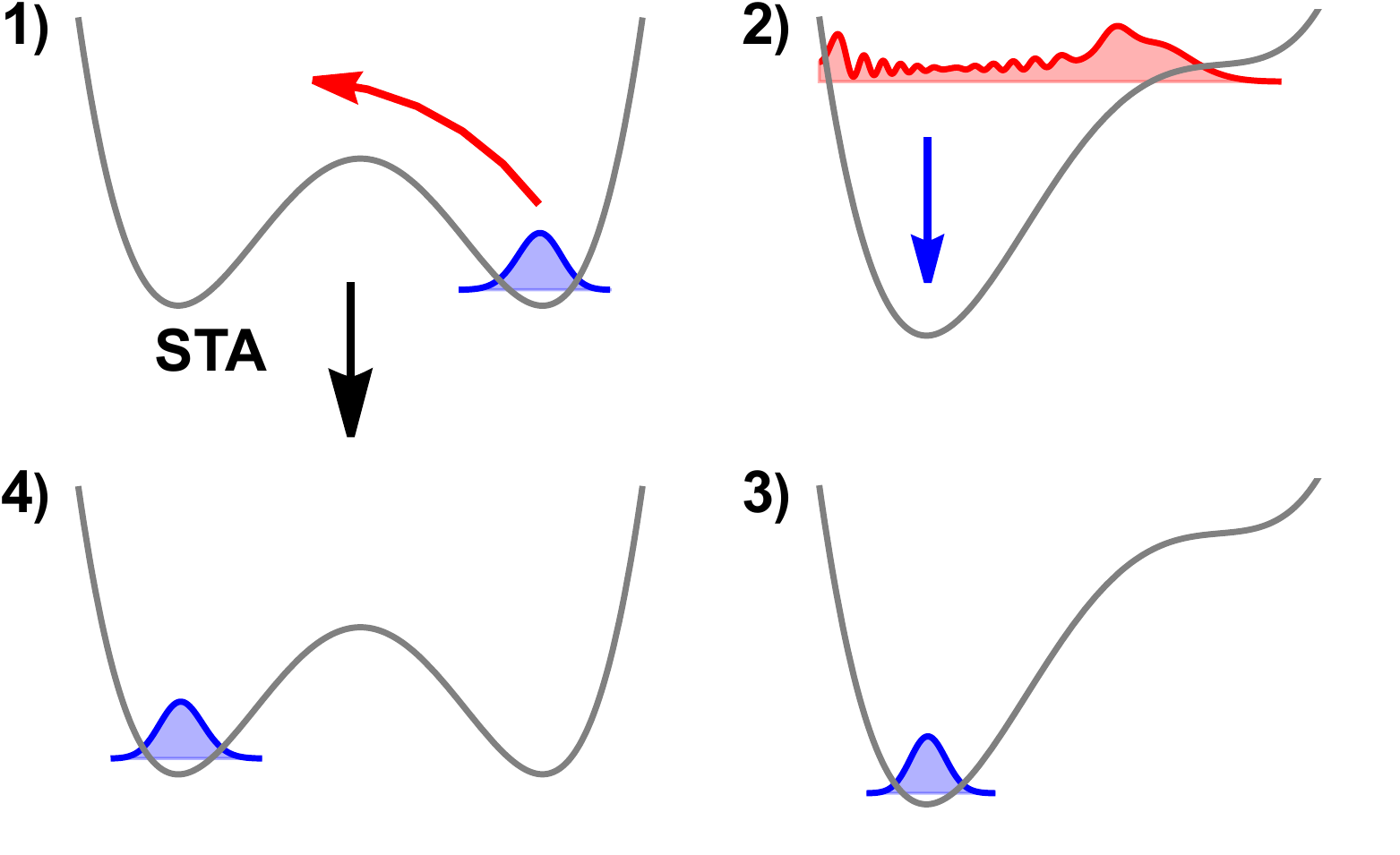}
    %\end{tabular}
    \end{center}
    \caption[example] 
    %>>>> use \label inside caption to get Fig. number with \ref{}
    {\label{fig3}
    Schematic illustration of the classical and the quantum state transfer protocols. The classical protocol follows the frame sequence $\mathbf{ 1)\to2)\to3)\to4)}$. The quantum protocol goes directly from $\mathbf{ 1)\to4)}$ with the help of STA. Figure taken from Ref.~\citenum{Qiongyuan_2023}.}
\end{figure}

As an application, we focus on the problem of transferring a specific state $\psi$ from the right to the left well in a double-well potential by considering different classical and quantum protocols. The Hamiltonian of the system is 
\begin{equation}\label{equ:doublewellsystemfree}
      H_\text{free} = \frac{p^2}{2m} + c_1 x^2 + c_2 x^4 ,
\end{equation}
where the coefficients $c_1<0$ and $ c_2>0$ determine the shape of the double-well potential, which corresponds to the black line of the frame 1) in \cref{fig3}.
    
The classical protocol is implemented by adding to the Hamiltonian in \cref{equ:doublewellsystemfree} a control function $f(x,t)$, which deforms the potential
\begin{equation}\label{eq.H0}
    H_0=H_\text{free}+f(x,t).
\end{equation}
For example, with $f(x,t)=\alpha^\text{\tiny C}(t)x$, one can tilt the potential as depicted in \cref{fig3} be choosing the proper function $\alpha^\text{\tiny C}(t)$. 
If performed quickly (i.e., non-adiabatically), the protocol  unavoidably excites the system to a high-energy state, as shown by the red
wavepacket in frame 2). Then, the system needs to be cooled back to the initial energy, as shown in frame 3). Finally, the potential can be brought back to the original form, as shown in frame 4). Faster is the protocol, larger is the energy increase and consequently the time required for the corresponding dissipation. This is eventually a situation one wants to avoid.
 \\

On the other hand, the quantum protocol is  implemented by adding to the modified Hamiltonian $H_0$ a counter-adiabatic term $H_\text{\tiny STA}$, which is defined as \cite{Berry_2009,Guery-Odelin_2019}
\begin{equation}
    H_\text{\tiny STA}(t) =i\hbar\sum_{i\neq j}\frac{\bra{i}\dot{H_0}\ket{j}}{E_j-E_i}\ket{i}\bra{j}\label{equ:cdterm}  ,
\end{equation}
and leads, under suitable assumptions, to a shortcut to adiabaticity (STA). Fundamentally, a state-transfer generated by the Hamiltonian $ H_0$ in Eq.~\eqref{eq.H0} can exploit the tunnelling effect in a double-well potential. However, a protocol constructed in such a manner would well only work in closed systems and in the adiabatic limit. Indeed, the interaction with the surrounding environment would have detrimental effect on the performance of the protocol. Conversely, the addition of STA Hamiltonian $H_\text{STA}$ helps the protocol to achieve such state-transfer in a finite time by accelerating the process. In such a way, the system can follow the adiabatic trajectories in the tunneling process but on a much shorter time-scale. \\

\begin{figure} [ht]
    \begin{center}
    \begin{tabular}{c c} %% tabular useful for creating an array of images 
    \includegraphics[width=6cm]{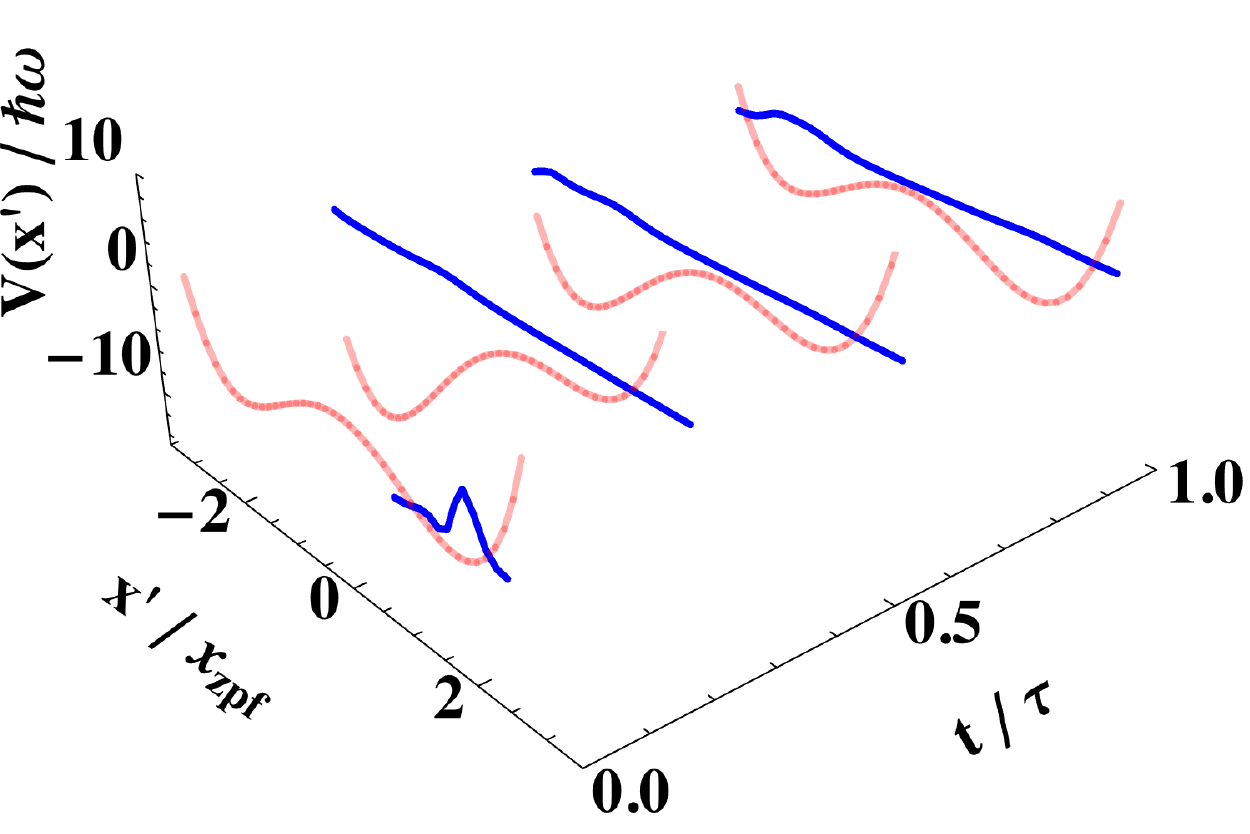}
    &
    \includegraphics[width=6cm]{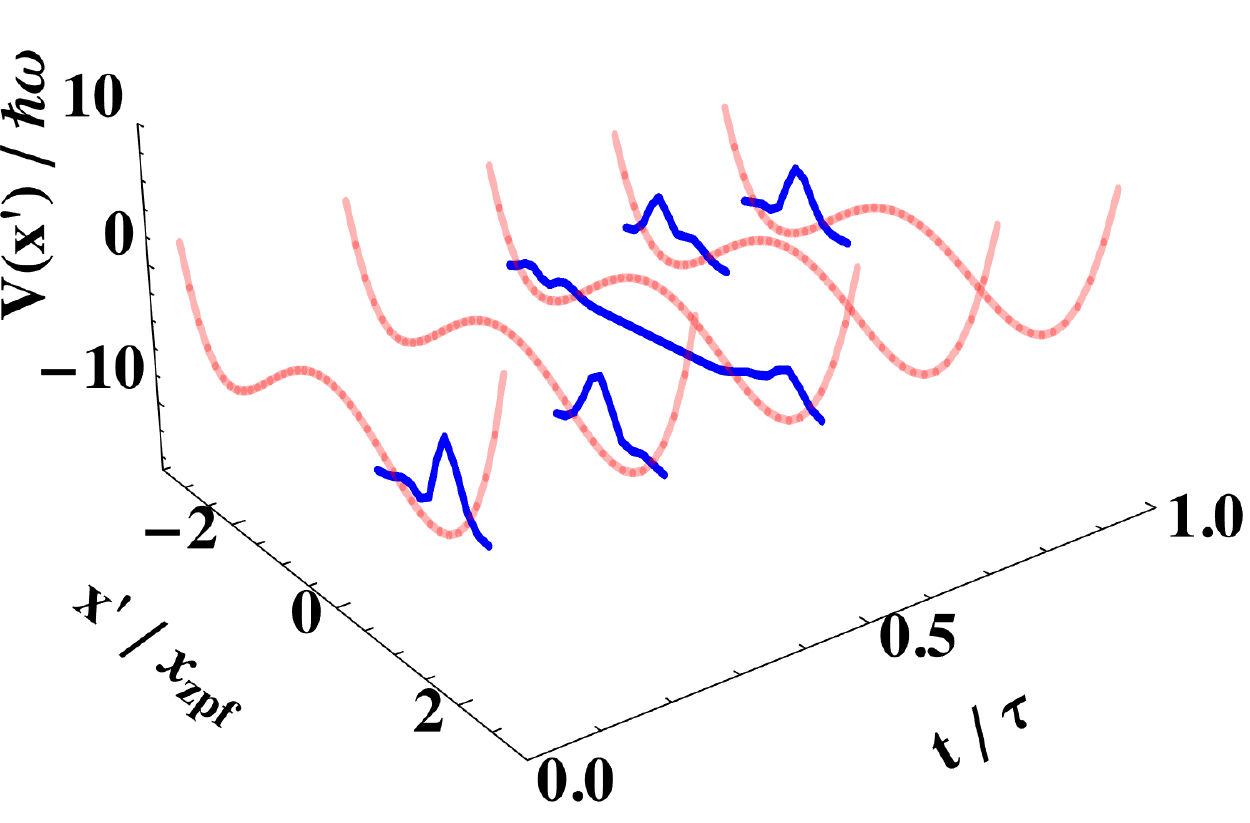}
    \end{tabular}
    \end{center}
    \caption[example]
    %>>>> use \label inside caption to get Fig. number with \ref{}
    {\label{fig4}
    Comparison of the classical (left panel) and the quantum (right panel) state transfer protocols at 1\,K. The blue lines represent the position density distribution $|\psi(x)|^2$ of the system as the potential (red line) is changed during the protocol. Figure taken from Ref.~\citenum{Qiongyuan_2023}.}
\end{figure}

The comparison of the classical and the quantum protocols in terms of the position distribution $|\psi(x)|^2$ is shown in \cref{fig4}, where we assumed a thermal environment of 1\,K. It is clearly visible, that the quantum protocol requires a smaller deformation of the potential, and it fully transfers the system from the right to the left well. Conversely, in the classical case, one has a strong deformation of the potential and the system ends in a state being spread over the entire potential. We also note that the time required for the classical protocol $\omega\tau=300$ is by far longer than that in the quantum case $\omega\tau=10$.\\

\begin{figure} [ht]
    \begin{center}
    %\begin{tabular}{c} %% tabular useful for creating an array of images 
    \includegraphics[width=8cm]{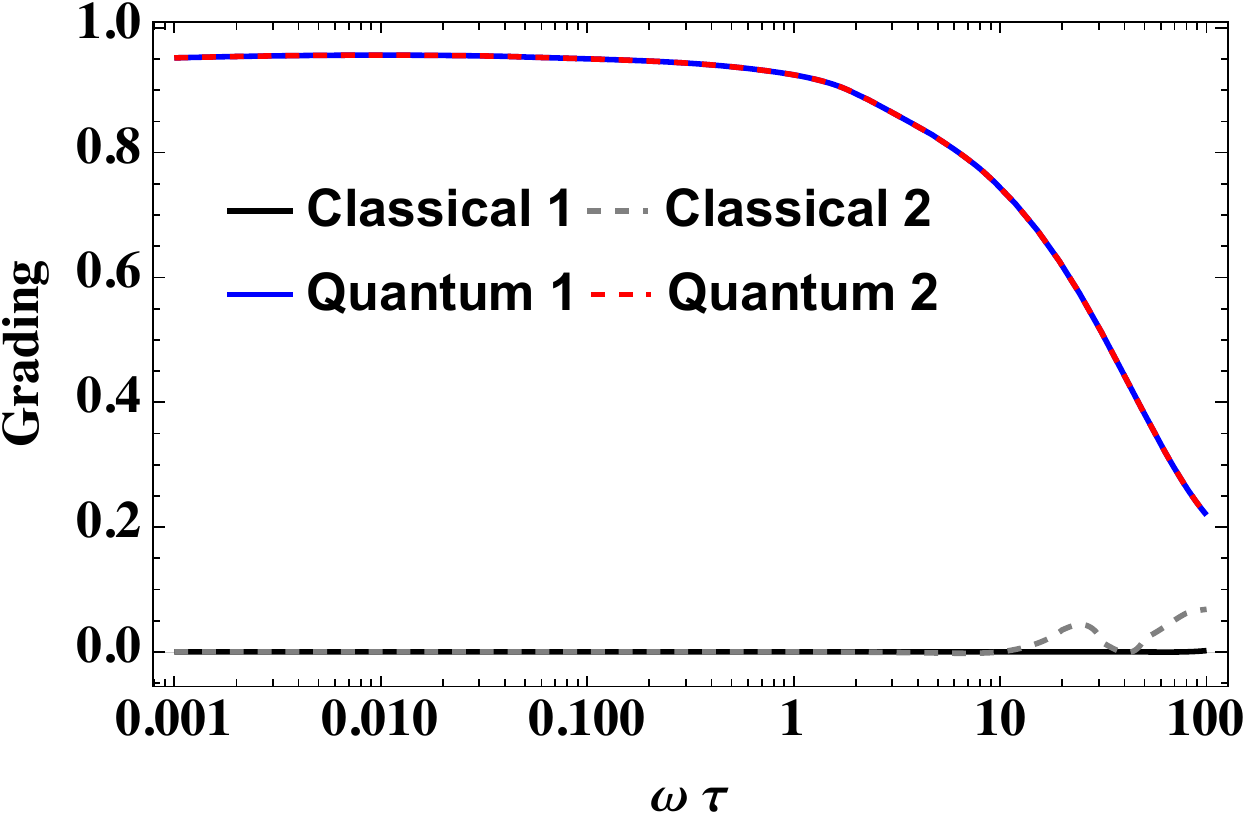}
    %\end{tabular}
    \end{center}
    \caption[example]
    %>>>> use \label inside caption to get Fig. number with \ref{}
    {\label{fig5}
    Protocol grading $\mQF$ for four state transfer protocols considered in cite{}, under the action of the environment at 1\,K. Figure taken from Ref.~\citenum{Qiongyuan_2023}.}
\end{figure}

Having introduced the classical and quantum protocols,  we now 
quantify their performances in terms of the protocol grading $\mQF$, which is shown in \cref{fig4}. For small values of $\omega\tau$, the protocol grading strongly favor the quantum protocol ($\mQF\sim 1$) over the classical one ($\mQF\sim 0$). Indeed, the quantum protocol well performs for protocol times $\tau$ that are smaller than the decoherence time imposed by the interaction with the surrounding environment. On the other hand, for larger values of $\omega\tau$, we approach a more adiabatic time-scale where also the classical protocol starts to perform better, although the quantum one is still outperforming the classical counterpart.

\acknowledgments % equivalent to \section*{ACKNOWLEDGMENTS}       

QW is supported by the Leverhulme Trust Research Project Grant (grant nr.~RGP-2018-266).
MC is supported by UK
EPSRC (Grant No.~EP/T028106/1).

\bibliography{report} 
\bibliographystyle{spiebib}

\end{document}